 \documentclass[final,5p,times,sort&compress,twocolumn]{elsarticle}

\usepackage{amssymb}
\usepackage[utf8]{inputenc}
\usepackage[T1]{fontenc}
\usepackage[hidelinks]{hyperref}
\DeclareUnicodeCharacter{2212}{-}
\usepackage{xcolor}
\usepackage[normalem]{ulem}

\journal{Physics Review C (ADP-25-10/T1272)}

\begin{document}

\begin{frontmatter}

\title{Heavy Neutron Star Phenomenology with an H-dibaryon}


\author[first]{Jesper Leong}
\ead{jesper.leong@adelaide.edu.au}
\author[first]{Pierre~A.~M.~Guichon}
\author[first]{Anthony W.~Thomas}

\affiliation[first]{organization={CSSM and ARC Centre of Excellence for Dark Matter Particle Physics, Department of Physics, University of Adelaide},
            postcode={5005}, 
            state={SA},
            country={Australia}}

\begin{abstract}
 The equation of state for dense nuclear matter in $\beta$-equilibrium is explored including the possibility of a doubly-strange H-particle. Consistent with experimental constraints, the mass of the H in free space is taken to be near the $\Lambda \, \Lambda$ threshold. Within the quark-meson coupling model, which we use, no new parameters are required to describe the interaction between the H-dibaryon and the other baryons. The maximum mass is only slightly reduced and the tidal deformability is essentially unchanged with this addition. In heavy neutron stars the H is abundant and extends as far as 6 km from the center of the core. 
\end{abstract}



\begin{keyword}
dense nuclear matter \sep neutron star \sep QMC \sep equation of state \sep hyperons \sep short range repulsion \sep excluded volume effect



\end{keyword}

\end{frontmatter}



\section{Introduction}
New observations by NICER and the gravitational wave detectors make this an exciting time to study neutron stars. Theoretical predictions which can be compared with these observations are hampered by our lack of understanding of matter at the extreme densities expected in these stars. Apart from the usual suspects, namely neutrons, protons, electrons and muons, one expects to find hyperons in the heavier stars and many authors have suggested other exotic particles may also appear; such as $\Delta$ baryons and dibaryons. There is also a possibility that the cores of the heavier stars may contain deconfined quarks and gluons.

Our focus is the potential role of an H-dibaryon.
This particle with quantum numbers spin and isospin zero, $B=2$ and $S=-2$, was first proposed in 1977 as a deeply bound, exotic 6-quark bag~\cite{Jaffe:1976yi}. 
This inspired a large number of experiments aimed at finding this particle, with negative results~\cite{Gal:2024nbr,Ejiri:1989rs,Dover:1988hg, Iijima:1992pp, BNLE836:1997dwi, Takahashi:2001nm, Yoon:2007aq}. A particularly interesting experimental search for the H involves the examination of double $\Lambda$ hyper-nuclei~\cite{Carames:2013hla, Sakai:1999qm}. Here a nucleus is formed with two $\Lambda$ hyperons in an emulsion. If the H were deeply bound, the double $\Lambda$ hypernucleus should decay into the original nucleus plus the H-dibaryon. A few double $\Lambda$ nuclei have been reported through $\Xi$ capture; $^{10}_{\Lambda\Lambda}Be$~\cite{Danysz:1963zz, Danysz:1963zza}, or interpreted as $^{13}_{\Lambda\Lambda}B$ \cite{Yamamoto:1991vf,Aoki:1991ip, Dover:1991kf}, and $^{6}_{\Lambda\Lambda}He$ \cite{Prowse:1966nz, Takahashi:2001nm}. No such decay of $\Lambda-\Lambda$ nuclei has been found, ruling out a deeply bound H. The NAGARA event at J-PARC ($^{6}_{\Lambda\Lambda}He$) yields the most stringent constraint, namely $M_H>2224$ 
MeV~\cite{Takahashi:2001nm}. 

While disappointing, the absence of a deeply bound H is consistent with the analysis of Mulders and Thomas~\cite{Mulders:1982da}, who showed that the larger size of the 6-quark bag would reduce the pion self-energy leading to a mass prediction around the mass of two $\Lambda$ hyperons. Uncertainties in that calculation meant that one could not be sure whether the H would be slightly bound or within a few 10s of MeV of that threshold.

Lattice QCD calculations from the HAL~\cite{INOUE201228} and NPLQCD~\cite{NPLQCD:2012mex} collaborations found a bound H close to the $\Lambda$-$\Lambda$ threshold but with heavy $u,d$ quarks. This motivated studies of the extrapolation to the physical $u$ and $d$ masses. For example, Shanahan \textit{et al.}~ \cite{Shanahan:2013yta} used a chiral extrapolation to show that if it is indeed an exotic 6-quark state the H is unbound by about $26 \pm 11$ MeV. If this is the case, one might expect that in dense matter in $\beta$-equilibrium the H might appear soon after the $\Lambda$ hyperon. Then the H might well exist inside the cores of heavier neutron stars (NS) where the central core density is many times the saturation density of normal nuclear matter~\cite{Tamagaki:1990mb}. 

The temperature of neutron stars is so low that it is natural to assume that it is below the critical temperature for Bose-Einstein condensation of the H particles. In the condensate all the particles are in the lowest state and therefore the kinetic energy is so low that it can barely resist to any attractive H-H interaction. The condensate then may collapse. However this instability is unlikely in the H system because, at the density where the H can appear, the attractive H-H interaction due to $\sigma$ exchange is overcompensated by the repulsive $\omega$ exchange~\cite{Faessler:1997mb}. Moreover in the quark-meson coupling (QMC) model that we use here the scalar polarisability effect reduces the $\sigma -H$ coupling at high density. This is one more support to the H condensate hypothesis.

The early calculations of the role of the H in NS generally worked with the hypothesis that it was bound. Furthermore, the strength of its coupling to the relativistic mean fields was not well understood. Typically, the $\omega$ coupling is fixed, which may be achieved by counting the number of non-strange quarks~\cite{Glendenning:1998wp}, while the strength of the $\sigma$ coupling was varied subject to the constraint that the H does not appear at ordinary saturation density. In one study the potential of the H at saturation was slightly attractive~\cite{Glendenning:1998wp}. This is in contrast to a more recent study which suggested that a more repulsive interaction is necessary for the H if one is to generate sufficiently heavy stars~\cite{Wu:2024vvw}.

Here we consider the role of the H-dibaryon using the quark-meson coupling model, which has been successfully applied to heavy NS, even when hyperons are included~\cite{Rikovska-Stone:2006gml, Motta:2020xsg, Leong:2023yma, Leong:2023lmw}. The uniqueness of QMC stems from the coupling of the mesons directly to the confined quarks, leading to important changes in the structure of the baryons in-medium~\cite{Guichon:2018uew,Guichon:1995ue}. While the model was originally based upon the MIT bag model~\cite{Chodos:1974pn}, similar features have been demonstrated within the NJL model~\cite{Bentz:2001vc,Mineo:2003vc,Cloet:2005rt}. Because it is the couplings of the mesons to the light quarks which are fundamental to this approach, the meson couplings to {\em all} hadrons are known for a given model of baryon structure {\em with no new parameters}. In particular, the meson couplings to the H-dibaryon are not free parameters. In contrast with the situation for $\Lambda$ hypernuclei, there have been very few observations of $\Xi$ hypernuclei. In any case, the QMC predictions of the binding energies of the known hypernuclei are in reasonable agreement with data~\cite{Hashimoto:2006aw,Nakazawa:2015joa,Shyam:2019laf,Guichon:2008zz}. The advantage of using QMC to model the H is that there is no \textit{a posteriori} fitting of the meson couplings of the H to fit NS observations. Either the H exists in NS and matches heavy NS observations, or it is not present in heavy NS.

In section 2 we summarize the additions to the published QMC formalism necessary to include the H-dibaryon. The results for the EoS and the properties of NS are described in section 3, while section 4 presents some concluding remarks.
 
\section{H-dibaryon in the QMC equation of state and Parameterisations}
A comprehensive review of the QMC model may be found in Refs.~\cite{Guichon:2018uew,Rikovska-Stone:2006gml}. The QMC EoS, and its parameterizations, are provided in Ref.~\cite{Leong:2023yma}. There we included a phenomenological ''overlap'' term to model possible short-distance repulsion at high densities not accounted for by the $\omega$ mean field.  

In this model the coupling of the baryon to the scalar mean field yields an effective mass which can be parameterized in terms of $\sigma$~\cite{Rikovska-Stone:2006gml} (see 
Eq.~(\ref{eq:effM}) ).
\begin{equation}
    \label{eq:effM}
    M^*_f(\sigma)=M_f-w_f g_\sigma \sigma + \tilde{w}_f \frac{d}{2} (g_\sigma \sigma)^2.
\end{equation}

The flavor dependent weightings, $w_f$ and $\tilde{w}_f$, depend on the choice of $R_N^{free}$, the MIT bag radius in free space. $R_N^{free}=1.0$ fm is used here \cite{Rikovska-Stone:2006gml}. The number of non-strange quarks in the H-dibaryon is twice that in the $\Lambda$. Since the scalar field in the QMC model couples only to the light quarks, the weightings $w_H$ and $\tilde{w_H}$ are approximatively twice that of the $\Lambda$ hyperon. The non-linear term in Eq.~(\ref{eq:effM}) is a feature of QMC related to the response of the internal structure of the baryon to the applied scalar field. It is the origin of the many body force, with its strength encoded by the scalar polarisability, $d$. 

This scalar polarizability is sensitive to $R_N^{free}$ but to a good approximation the bag radius is the same for all baryons in the octet. However, because of the increase in the number of confined quarks in the H-dibaryon, Mulders and Thomas found an increase by around 20\% in the free bag radius for the H particle~\cite{Mulders:1982da}. Here we take $R_H=1.2 \times R_N^{free}$. Thus the in-medium mass of the H-particle is
\begin{eqnarray}
    M^*_H&=M_H-2\times[0.6672+0.0462 R_H-0.0021 {R_H}^2] g_\sigma \sigma \nonumber \\
    &+2\times[0.0016+0.0686R_H-0.0084 {R_H}^2](g_\sigma \sigma)^2 \, . 
    \label{eq:effMH}
\end{eqnarray}
\begin{figure*}
    \centering
    \includegraphics[scale=1]{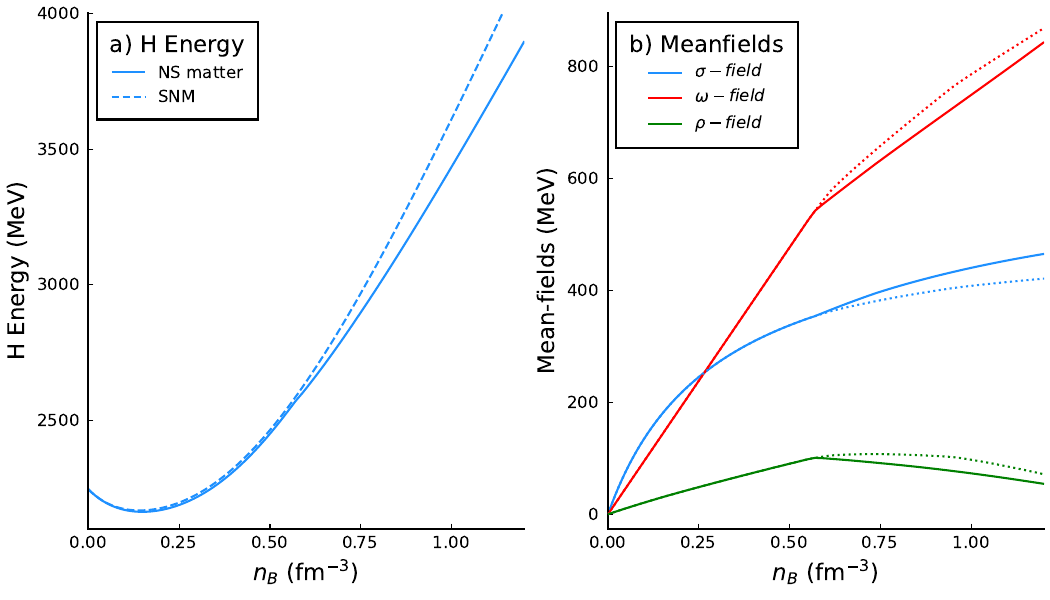}
    \caption{In a) the energy of the H particle in SNM (solid) and NS matter  at $\beta$-equilibrium (dash) are presented. In b) the mean-fields with nucleon couplings in NS matter are shown with (solid) and without (dot) the H dibaryon. Since the neutron has isospin of $-1/2$, the isovector $\rho$ is inverted to present positive values. Both results in this figure are for $M_H=2247$ MeV.}
    \label{fig:Henergy_meanfields}
\end{figure*}
\begin{figure}
    \centering
    \includegraphics[width=1.0\linewidth]{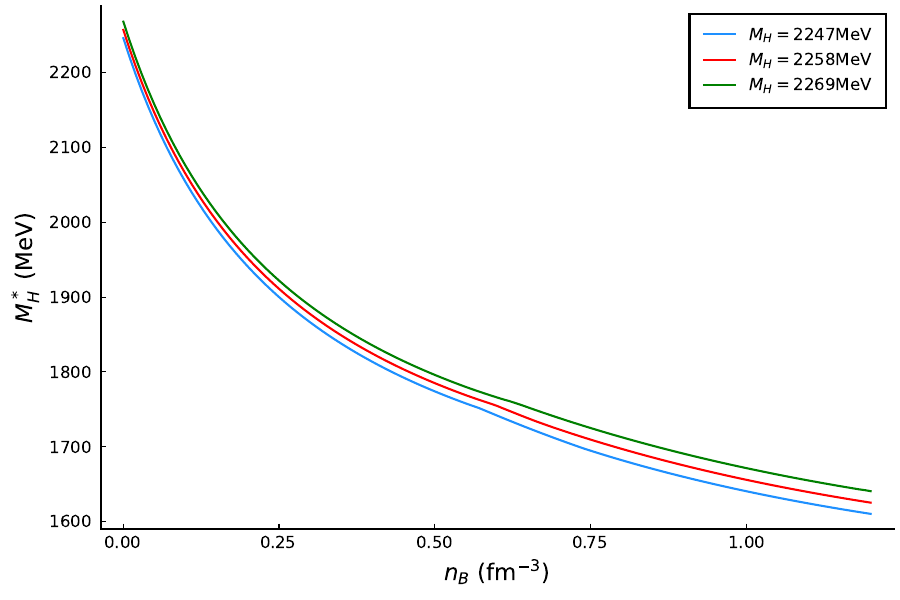}
    \caption{The effective mass of the H as given by Eq.~(\ref{eq:effMH}).}
    \label{fig:HeffM}
\end{figure}
\begin{figure*}
    \centering
    \includegraphics[width=1.0\linewidth]{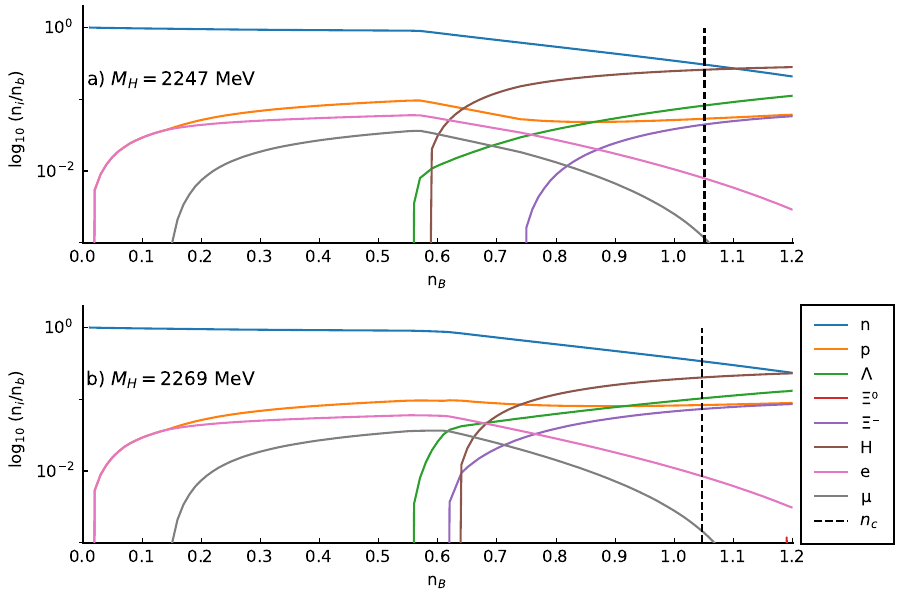}
    \caption{Species fractions for $\beta$-stable matter. In a) $M_H= 2247$ MeV, and b) $M_H= 2269$ MeV, we show the particle fractions as a function of total number density. $n_c$ denotes the central number density for the maximum mass NS as depicted by the black dashed vertical line with the precise values listed in Table \ref{tab:central}.}
    \label{fig:species}
\end{figure*}
\begin{figure*}
    \centering
    \includegraphics[width=1.0\linewidth]{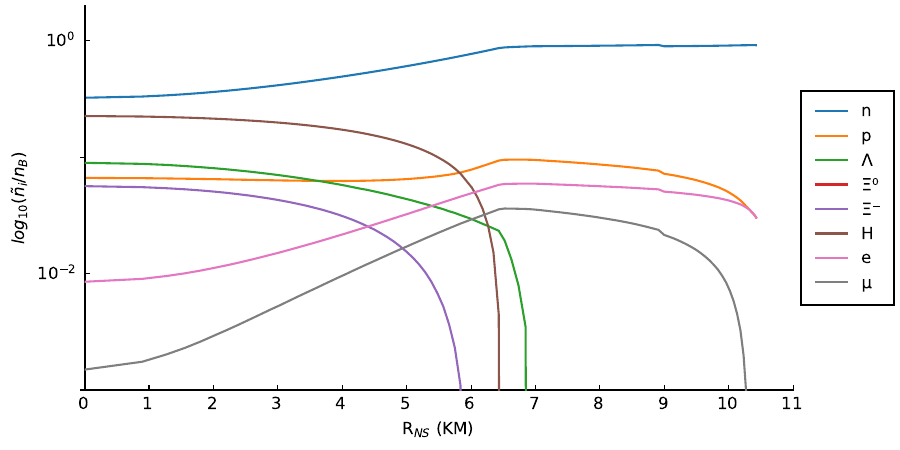}
    \caption{Particle population within the interior core of a maximum mass NS when $M_H=2258$ MeV. The species fractions is truncated at the crust-core boundary.}
    \label{fig:radspecies}
\end{figure*}
\begin{figure*}
    \centering
    \includegraphics[width=1\linewidth]{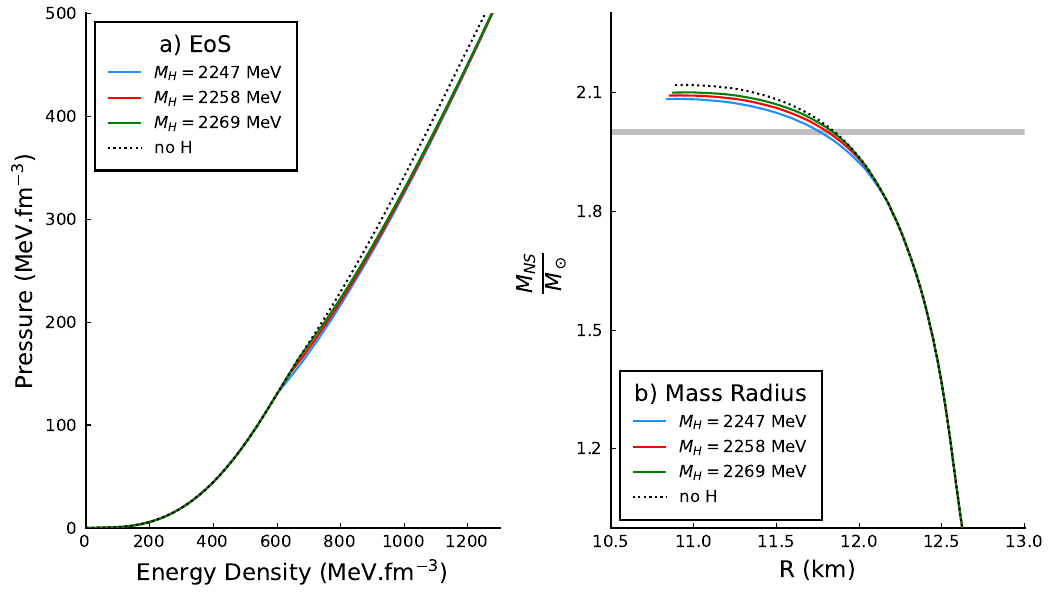}
    \caption{a) The QMC EoS for difference masses of H. In the dotted black line we show the EoS without the H-dibaryon~\cite{Leong:2023yma}. The curves begin to deviate once the H appears. b) The resultant mass-radius curves calculated using the TOV equation. The gray horizontal line indicates a $2.0$ $M_\odot$ star.}
    \label{fig:EoS_MR}
\end{figure*}

The only free parameter when incorporating the H into QMC is therefore its free mass, $M_H$. As discussed in the previous section, this mass is currently unknown. The best experimental lower bound suggests that it cannot be bound by more than about ten MeV with respect to the $\Lambda-\Lambda$ threshold. $M_H$ is given in Eq.~(\ref{eq:freeMH}).
\begin{equation}
    \label{eq:freeMH}
    M_H=2 M_\Lambda-B_H \, .
\end{equation}
We presume that the H is unbound and use as guidance the result from Shanahan \textit{et al.} that $B_H=-26\pm11$ MeV~\cite{Shanahan:2013yta}.

The Lagrangian density for the H-dibaryon is
\begin{equation}
\label{eq:Hlag}
    \mathcal{L}_D=\mathcal{D}^*_\mu H^* \mathcal{D}^\mu H - {M^*_H}^2 H^*H,
\end{equation}
where $\mathcal{D}_\mu=\partial_\mu+ig_{\omega,H} \omega_\mu$ includes the coupling to the $\omega$ vector field \cite{Glendenning:1998wp, Faessler:1997mb}. Since H is a boson with zero isospin the $\rho$ does not couple to it. 

The Euler-Lagrange equation is then
\begin{equation}
    \label{eq:ELH}
    ( \mathcal{D}_\mu \mathcal{D}^\mu+{M^*_H}^2) H=0 \, .
\end{equation}
The H particle is a Boson and hence forms a condensate with $\vec{k}=0$. At mean-field level one can show that for an s-wave condensate the chemical potential is
\begin{equation}
    \label{eq:HH}
    \mu_H=M^*_H+g_{\omega,H}\omega \, . \label{eq:chemH}
\end{equation}
It follows that the density is simply
\begin{equation}
    \label{eq:Hdensity}
    n_H=2{M^*_H} H^* H \, .
\end{equation}

The H modifies the mean-fields as follows
\begin{eqnarray}
    \label{eq:sigmafield}
    {m_\sigma}^2\sigma+\lambda_3 \frac{{g_\sigma}^3}{2}\sigma^2=\sum_f n^s_f-\frac{\partial M^*_H}{\partial \sigma} n_H, \\
    \label{eq:omegafield}
    {m_\omega}^2 \omega = \sum_f g_{\omega,f} n^v_f + g_{\omega,H} n_H \, .
\end{eqnarray}
Here $\lambda_3=0.02$ $fm^{-1}$ is the coefficient of the self-interaction of the $\sigma$ field, which is needed to reproduce the giant monopole resonance energies in finite nuclei~\cite{Martinez:2018xep}. $n^s$ and $n^v$ are the scalar and vector densities, respectively, and the summation is over $n$, $p$, $\Lambda$, $\Xi^0$ and $\Xi^-$. The last terms in Eqs.~(\ref{eq:sigmafield}) and (\ref{eq:omegafield}) is the contributions of the H to the respective mean-fields~\cite{Glendenning:1998wp}. Finally, the H contribution to the energy is
\begin{equation}
    \label{eq:energyH}
    \epsilon_H=2 {M_H^*}^2 H^*H= M^*_H n_H \, .
\end{equation}

We note that the theory outlined above is  thermodynamically consistent~\cite{Faessler:1996ta, Faessler:1997mb, Glendenning:1998wp}. In Ref.~\cite{Glendenning:1998wp} the authors observed that there is no direct contribution from the H to the pressure. However, the pressure does change indirectly because of the modifications of the mean-fields, as we can see from Eqs.~(\ref{eq:sigmafield}) and (\ref{eq:omegafield}). 

Here we account for the potential additional short-distance repulsion that may be present at high densities by incorporating the overlap term used in Ref.~\cite{Leong:2023yma}. This is dependent on the total number density, $n_B=\sum_f n_f+2n_H$. In this case the H does change the pressure through both the modification of the mean-fields and the overlap mechanism. Tamagaki suggested that the hard-core radius of the H-dibaryon should be between $0.5-0.7$ fm~\cite{Tamagaki:1990mb}. The range parameter in the overlap model used here is consistent with this. For simplicity we presume that the H has the same hard-core radius as the rest of the baryons. 

\section{Results}
In Fig.~\ref{fig:Henergy_meanfields} a) the energy of the H dibaryon is presented in the case of symmetric nuclear matter and NS matter in $\beta$-equilibrium. The minimum energy of the H is found close to the saturation density of symmetric nuclear matter (SNM). The H is bound by $85$ MeV in NS matter and $79$ MeV in SNM. The difference is accounted for by the small change in the $\sigma$-field associated with the additional neutrons in NS matter (see Eq.~(\ref{eq:sigmafield}) ). 
For reference, the $\Lambda$ is bound by around $38$ MeV. The solid line representing the energy of the H in matter in $\beta$-equilibrium, as shown in 
Fig.~\ref{fig:Henergy_meanfields} a), starts to diverge from SNM when the H appears because of the modifications of the mean-fields. 
\begin{table}[]
    \centering
    \begin{tabular}{c|ccccc}
        $M_H$  & $M_{NS}$  & R & $n_c$  & $\varepsilon_c$ & $P_c$  \\ \hline
        - & 2.12 & 11.0 & 1.03 & 537 & 1314\\
        2247 & 2.08 & 10.91 & 1.05 & 537 & 1332 \\
        2258 & 2.09 & 10.91 & 1.05 & 537 & 1332 \\
        2269 & 2.10 & 10.94 & 1.05 & 537 & 1330 \\
    \end{tabular}
    \caption{The maximum mass ($M_{NS}$, $M_\odot$) and corresponding radius ($R$, km), for each choice of $M_H$ (MeV) is presented along with the central number density ($n_c$, fm$^{-3}$), energy density ($\varepsilon$, MeV-fm$^{-3}$) and pressure ($P_c$, MeV-fm$^{-3}$). For reference, the case with no H-dibaryon is presented at the top of the table~\cite{Leong:2023yma} \, .}
    \label{tab:central}
\end{table}

Figure~\ref{fig:Henergy_meanfields} b) shows the magnitudes of the mean fields experienced by nucleons in NS matter. The H-particle appears at around $0.6$ $fm^{-3}$ (see Fig.~\ref{fig:species} a)) and is subject to an overall repulsive interaction which allows the H condensate to be stable. After the H appears the strength of the $\omega$-field is weakened, whilst the strength of the $\sigma$ is increased. This is primarily caused by the changes in neutron density, as can be seen in the weakening of the iso-vector vector mean field (carried by the $\rho$-meson), to which the H does not couple. In Fig.~\ref{fig:HeffM} we show the effective mass of the H, $M_H^*$, which is reduced in response to the scalar attraction.

The H-particle is electrically neutral but carries baryon number 2. Thus, the only modification to $\beta$-equilibrium is to the conservation of baryon number. Figure~\ref{fig:species} shows the species fractions for a) $M_H=2247$ MeV, and b) $M_H=2269$ MeV. The H particle always appears after the $\Lambda$, given that we have chosen $M_H$ such that H lies above the $\Lambda-\Lambda$ threshold. The threshold density of the H is increased with increasing $M_H$. The H is in competition with the hyperons and besides a reduction in the abundance of the other baryons, the presence of the H increases the threshold density at which any new hyperon appears after the H. In this case, those hyperons are the $\Xi^{-}$ and $\Xi^{0}$. Without the H, the $\Xi^0$ baryons were present in the QMC EoS at high densities~\cite{Leong:2023yma} but once the H is included the $\Xi^0$ does not appear at all. In Fig.~\ref{fig:species} the central density for the maximum mass star is indicated by the black dashed line for $M_H= 2247$ MeV and $2269$ MeV. Near this threshold the H is abundant, indicating that the cores of heavy NS may contain a sizable fraction of H-matter condensate. Figure~\ref{fig:radspecies} depicts the population of baryons and the H-dibaryon in the interior of a maximum mass star for $M_H=2258$ MeV. The maximum mass, radius and central energy density, pressure and number density is located in table \ref{tab:central}. The H particle is highly abundant in the center of the star and extends out to a radius of around 6.5 km.

The EoS is presented in Fig.~\ref{fig:EoS_MR} a). The low density EoS presented by Hempel and Schaffner-Bielich was used to accurately model the crustal regions of the NS~\cite{Hempel:2009mc, Hempel:2011mk}. The EoS for the case with no H is indicated with the dotted black line. As expected, the appearance of the hyperon and the H particle softens the EoS as each of their threshold densities are reached. The overlap interaction mitigates the softening of the EoS normally associated with hyperons due to the loss of neutron degeneracy pressure. The same is also true of the H. 

As we see in Fig.~\ref{fig:EoS_MR} b), the presence of the H does slightly reduce the maximum mass predicted by the QMC model. Nevertheless, the maximum masses obtained with all 3 choices of $M_H$ exceed $M_{NS}>2.0$ $M_\odot$. Although the choice of $M_H$ changes the density at which the H-dibaryon appears, there is no great reduction in the maximum mass of the NS, with a larger value of $M_H$ yielding a smaller reduction in the maximum mass. Also note that $M_{NS}=1.4$ $M_\odot$ does not contain any hyperons or H, and therefore the tidal deformability for $\Lambda_{1.4}=430$ remains the same as previously reported by Leong 
{\em et al.}~\cite{Leong:2023yma}. In Table~\ref{tab:central} we display the central properties for no H and each choice of $M_H$ at maximum mass. The central values of density, energy and pressure are all similar. 

\section{Conclusion}
\label{sec:conclusion}
We have examined the effects on the EoS of dense matter in $\beta$-equilibrium when one includes the possibility of an H-dibaryon with mass somewhat higher than the $\Lambda-\Lambda$ threshold. The QMC model was chosen because all the meson couplings to the H are predicted within the model, rather than being free parameters. 

The TOV equation was then solved to obtain the masses and radii of neutron stars generated by the new EoS. While the presence of the H-dibaryon does tend to lower the central density and pressure for the heaviest stars, the maximum masses remain above 2 $M_\odot$. In addition, because the H-dibaryon only appears well above three times the saturation density of nuclear matter, the successful predictions for tidal deformability reported in earlier work remain unchanged. 

The results presented here show that even though the H-dibaryon is predicted to be bound at normal nuclear densities, its presence is still compatible with observations of heavy NS and the observed tidal deformability.

\section*{Data availability}

Data will be made available upon request.

\section*{Acknowledgements}
This work was supported by the University of Adelaide and the Australian Research Council through a grant to the ARC Centre of Excellence for Dark Matter Particle Physics (CE200100008) and DP230101791 (AWT).

\bibliographystyle{elsarticle-num} 
\bibliography{apssamp}






\end{document}